\documentclass[twocolumn,showpacs,amsmath,prl,aps,amssymb,superscriptaddress,nofootinbib]{revtex4-1} 
\usepackage[T1]{fontenc}
\usepackage[latin9]{inputenc}
\usepackage{times}
\usepackage{color} 
\usepackage{xspace}
\usepackage{amssymb,amsmath}
\usepackage{amsbsy}
\usepackage[pdftex]{graphicx}
\usepackage{bm}
\usepackage{float}
\usepackage{adjustbox}
\usepackage{epstopdf}

\usepackage[unicode,breaklinks]{hyperref}
\hypersetup{
    unicode=true,
    plainpages=false, 
    colorlinks=true,
    linkcolor=blue,
    citecolor=blue,
    filecolor=black,
    urlcolor=blue
}
\usepackage{url}
\usepackage{verbatim}

\begin{document}

\title{Universal coarsening dynamics of a quenched ferromagnetic spin-1 condensate}
\author{Lewis A. Williamson}  
\affiliation{Dodd-Walls Centre for Photonic and Quantum Technologies, Department of Physics, University of Otago, Dunedin, New Zealand}
\author{P.~B.~Blakie}  
\affiliation{Dodd-Walls Centre for Photonic and Quantum Technologies, Department of Physics, University of Otago, Dunedin, New Zealand}

\begin{abstract} 
We demonstrate that a quasi-two-dimensional spin-1 condensate quenched to a ferromagnetic phase undergoes universal coarsening  in its late time dynamics.
The quench can be implemented by a sudden change in the applied magnetic field and, depending on the final value, the ferromagnetic phase has easy-axis (Ising) or easy-plane (XY) symmetry, with different dynamical critical exponents.  Our results for the easy-plane phase reveal a fractal domain structure and the crucial role of polar-core spin vortices in the coarsening dynamics.   
\end{abstract}

\maketitle

Ultra-cold atomic systems are well isolated from the environment and present a pristine system for exploring non-dissipative manybody dynamics. An emerging area of exploration with these systems involves the dynamics induced by a quench across a phase transition to a symmetry-broken phase. Following the quench domains form, with each of these domains having made an independent choice for the symmetry-breaking order parameter. An aspect that has seen experimental investigation  \cite{Weiler2008a,Navon2015a,Chomaz2015a,lamporesi2013} involves quantifying the production of topological defects that emerge between domains immediately after the quench \cite{Kibble1976a,Zurek1985a}.  
 Another aspect involves how these domains coarsen over time as the different broken-symmetry phases compete to select the equilibrium state.
  Often at late times, when the domains are large compared to microscopic length scales, the coarsening dynamics is universal: correlation functions of the order parameter collapse to a universal scaling function when the spatial coordinates are scaled by a characteristic length $L(t)$, where $t$ is the time after the quench \cite{Bray1994a}. The time dependence of this length scale $L(t)\sim t^{1/z}$ yields the dynamical critical exponent $z$.  Most work in the classical theories of phase ordering kinetics has focussed on dissipative models relevant to temperature quenches.  The late-time dynamics for systems undergoing conservative Hamiltonian evolution has developed as a new area of interest, particularly due to developments with ultra-cold atomic gases \cite{Damle1996a,zheng2000,koo2006,asad2007,Takeuchi2012a,Hofmann2014a}.
 
 Spinor Bose-Einstein condensates  exhibit both superfluid and magnetic order \cite{Kawaguchi2012R,StamperKurn2013a} and present a rich phase diagram \cite{Stenger1998a,Ho1998a,Ohmi1998a} for considering transitions between phases with different symmetry properties. The simplest non-trivial case is a spin-1 condensate, which has been realised using $^{87}$Rb and $^{23}$Na atoms with ferromagnetic and anti-ferromagnetic interactions, respectively. In general, an external magnetic field breaks the full spin symmetry of the Hamiltonian, reducing it to axial symmetry transverse to the field. The external field also leads to a quadratic Zeeman shift of the spin states that competes with the spin-dependent interaction to determine the equilibrium phase \cite{Stenger1998a}. This system is ideally suited to studying phase transition dynamics because the Zeeman energy can be dynamically varied in experiments, allowing quenches between phases \cite{Sadler2006a,Bookjans2011b} (also see  \cite{De2014a}), and because the subsequent magnetisation dynamics can be revealed with \textit{in situ} imaging \cite{Higbie2005a}.  While the short-time dynamics following the quench is well-understood (e.g.~see \cite{Lamacraft2007a,Saito2007a,Saito2007b,Uhlmann2007a,Damski2007a}),  the  subsequent domain coarsening   has been the subject of considerable debate \cite{Mukerjee2007a,Lamacraft2007a,Guzman2011a} and has been identified as a significant outstanding problem in the field \cite{StamperKurn2013a}.

\begin{figure}[htbp] 
   \centering
   \includegraphics[width=3.4in]{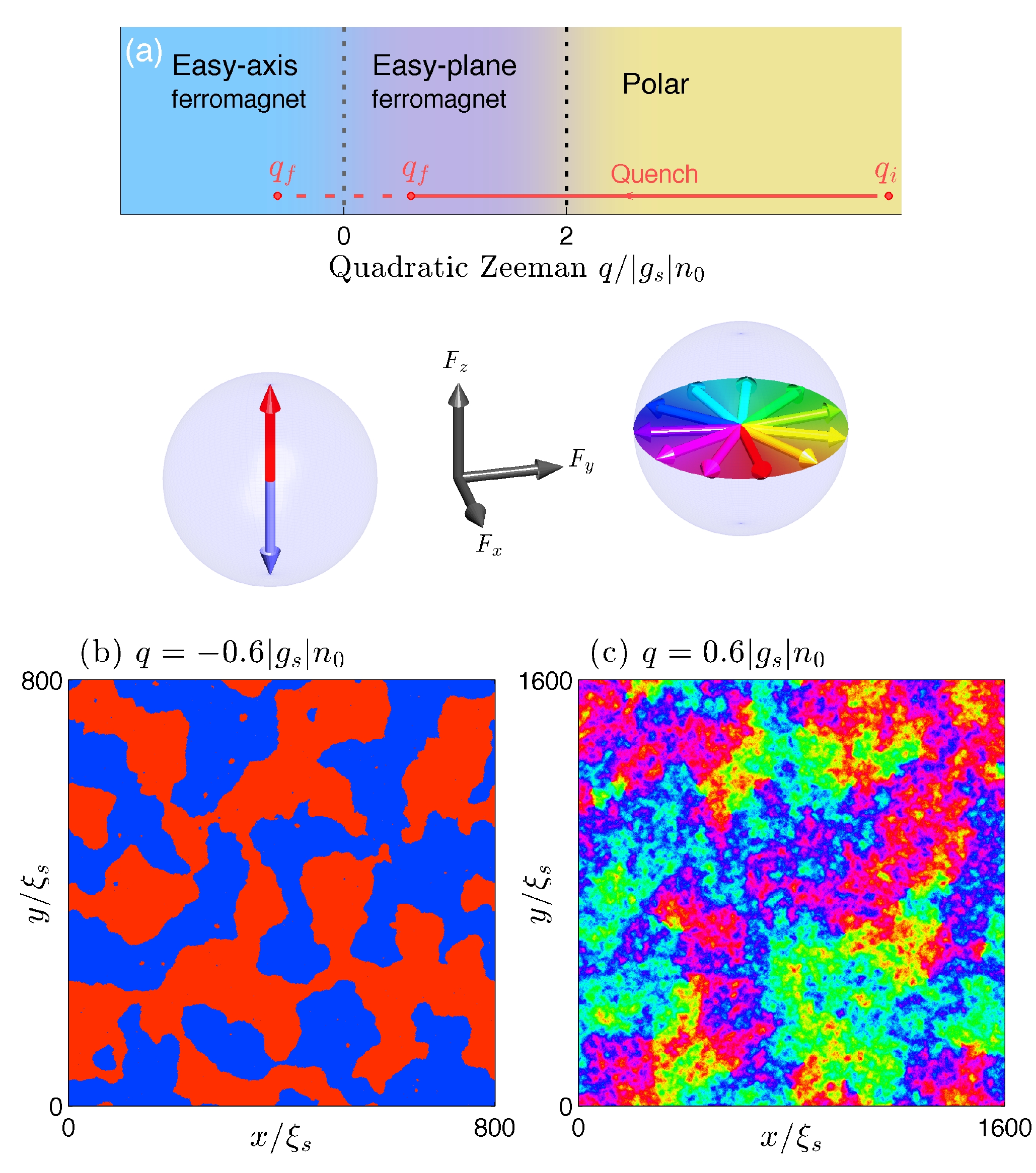} \\
   \vspace*{-0.4cm}
   \caption{(a) Phase diagram of a spin-1 BEC with $g_s<0$ and zero $z$-component of magnetisation indicating the transition between polar and ferromagnetic phases as the quadratic Zeeman energy $q$ is varied. The quenches of $q$ from $q_i$ to $q_f$ considered in this paper are indicated schematically. (b), (c) show the direction of the order parameter during late-time spin ordering for quenches into the easy-axis (b) and easy-plane (c) phases, with the color map indicated on the respective spin-spheres. Simulation parameters: $n_0=10^4/\xi_s^2$, $g_n/|g_s|=3$. }
   \label{fig1}
\end{figure}

Here we study the non-dissipative dynamics of a quasi-two-dimensional ferromagnetic spin-1 condensate. We consider the quantum phase transition of this system from an unmagnetised polar state to either an easy-axis or  easy-plane ferromagnetic state, depending on the final value of the quadratic Zeeman energy (see Fig.~\ref{fig1}). 
We demonstrate that both quenches behave universally in their late time coarsening dynamics. We find $1/z=0.68$ for the easy-axis phase, consistent with a binary fluid interpretation \cite{Furukawa1985a} and disagreeing with an earlier result of $1/z\approx1/3$ \cite{Mukerjee2007a}. A hydrodynamic analysis \cite{Kudo2013a} also obtained a $t^{2/3}$ growth law, and showed that the growth reduces to $t^{1/3}$  if the effects of superfluid flow are removed. A recent study of the coarsening dynamics of an immiscible binary condensate revealed a $t^{2/3}$ growth law and verified the scaling hypothesis by demonstrating correlation function collapse \cite{Hofmann2014a}.
 For the easy-plane case, we show that topological defects, namely polar-core spin vortices, play a crucial role in the dynamics, and find an exponent (accounting for logarithmic corrections to the coarsening) of $z=1.04$, consistent with the $z=1$ result for model E. An analysis of the structure factor for the quenches reveals an expected universal scaling with a $k^{-3}$ Porod tail for the easy-axis case, yet reveals a non-integer tail in the easy-plane case. We verify that this arises from a fractal structure of the domains. Such fractal behaviour in the Porod tails has also been observed in the dynamical scaling of aggregates in dense colloidal solutions  \cite{Sintes1994a}. Our observation of coarsening in the easy-plane phase demonstrates that the system continually anneals towards an equilibrium state (c.f. \cite{Barnett2011a}).

The  energy functional for a quasi-2D spin-1 condensate is \cite{Barnett2011a}  
\begin{align}
\!E\!=\!&\int \!d^2\bm{x}\,\!\left[\bm{\psi}^\dagger\frac{\hat{p}^2}{2M}\bm{\psi}+\frac{g_n}{2} n^2+\frac{g_s}{2}|\bm{F}|^2+q\bm{\psi}^\dagger f_z^2\bm{\psi}\right],\label{EqEfun}
\end{align} 
where $\bm{\psi}=\left(\psi_1,\psi_0,\psi_{-1}\right)^T$.  The system has density interactions $g_n n^2$  ($n=\bm{\psi}^\dagger\bm{\psi}$ is the total areal density) and  spin interactions $g_s |\bm{F}|^2$ ($\bm{F}$ is the spin density with components $F_\mu=\bm{\psi}^\dagger f_\mu\bm{\psi}$, where $f_\mu\in\{f_x,f_y,f_z\}$ are the spin-1 matrices). A magnetic field along $z$  shifts the energies of the spin states. The linear Zeeman shift has been removed by transforming $\bm{\psi}$ into a frame rotating at the Larmor frequency. The quadratic Zeeman shift  $q$ can be tuned independently of the magnetic field using external microwave fields (e.g.~see \cite{Gerbier2006a}).

For ferromagnetic interactions ($g_s<0$) the ground state of Eq.~(\ref{EqEfun}) can exist in three phases  dependent on the relative values of $q$ and $ng_s$ \cite{Kawaguchi2012R}. The phase diagram for a system with zero magnetisation along $z$ is shown in Fig.~\ref{fig1}(a), and has been explored in experiments with $^{87}$Rb \cite{Sadler2006a,Guzman2011a}.
Here we consider the coarsening dynamics of this system quenched from a unmagnetized \textit{polar} phase to a magnetized \textit{ferromagnetic} phase by a sudden change in $q$. For $0<q<2|g_s|n_0$, where $n_0$ is the initial condensate density, the magnetisation lies in the $xy$-plane (easy-plane) and we take $\phi=(F_x,F_y)/n_0\equiv\bm{F}_\perp/n_0$ as the order parameter. This phase breaks the continuous axial symmetry of the Hamiltonian, and the order parameter is not conserved. For $q<0$ the magnetization lies along the $z$ axis (easy-axis) and we take $\phi=F_z/n_0$ as the order parameter. This phase breaks the $\mathbb{Z}_2$ symmetry of the Hamiltonian, but in this case the order parameter is conserved.
 
 \begin{figure}[htbp] 
   \centering
   \includegraphics[width=3.5in]{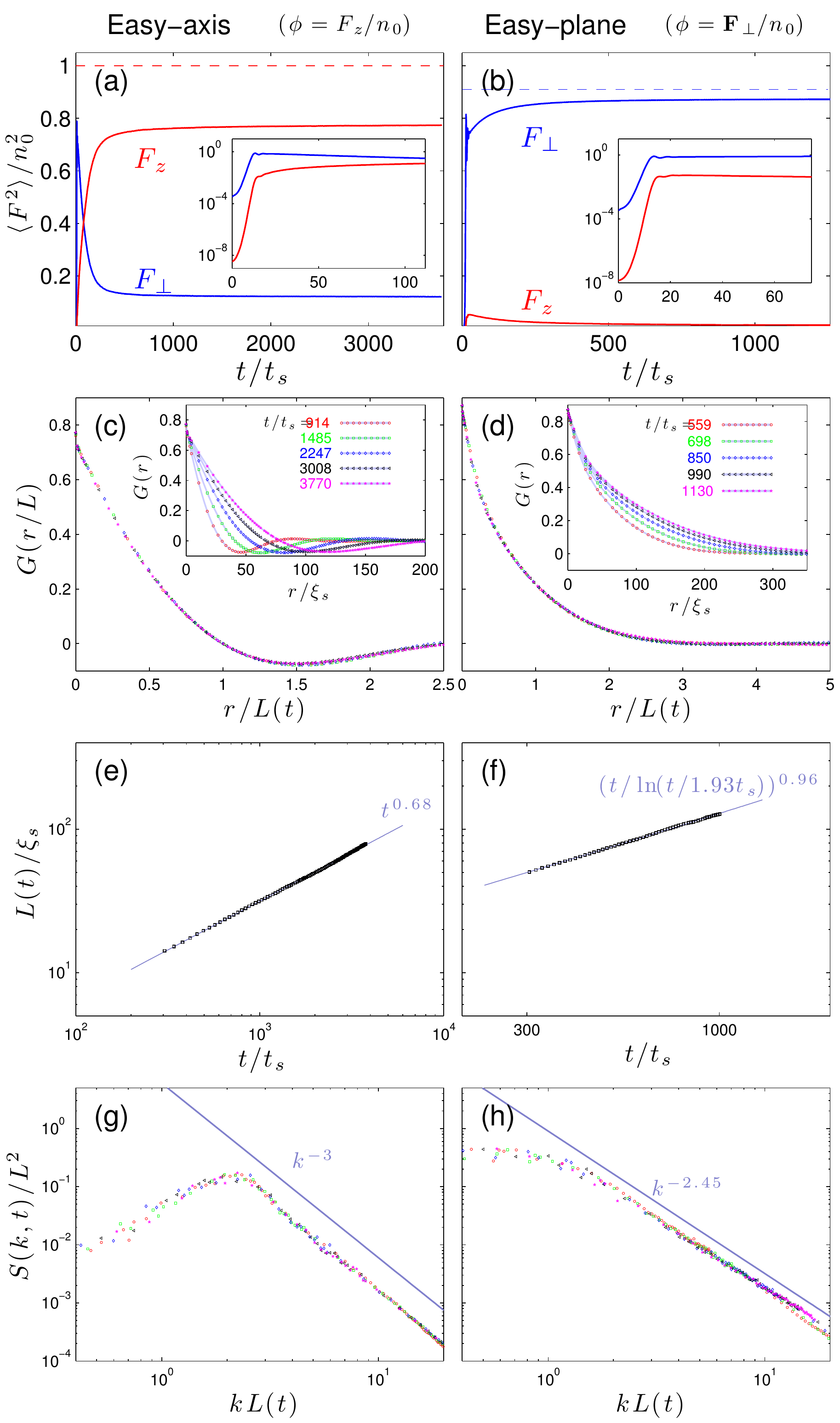} \\
   \vspace*{-0.25cm}
   \caption{Results for easy-axis (left column) and easy-plane (right column) coarsening dynamics. (a,b) Growth in longitudinal and transverse magnetisation. Dashed lines indicate ground state longitudinal (transverse) magnetisation for the easy-axis (easy-plane) cases. The excess energy per particle available for thermalisation is $Q_{\mathrm{EA}}=(\frac{1}{4}q_0-q_f)$, $Q_{\mathrm{EP}}=\frac{1}{4}q_0(1-q_f/q_0)^2$, for the easy-axis and easy-plane quenches respectively, where $q_0=2|g_s|n_0$. For our parameters $Q_{\mathrm{EA}}\approx4.5\times Q_{\mathrm{EP}}$, explaining greater  thermal depletion in the easy-axis quench \footnote{The initial noise in the condensate contributes an energy of approximately $3\!\times\!10^{-3}Q_{\mathrm{EA}}$ and $2\!\times\!10^{-2}Q_{\mathrm{EP}}$ in the EA and EP quenches, respectively.}. (c,d) Scaled order parameter correlation functions showing collapse, with unscaled data shown in the Insets. (e,f) Characteristic length scale $L(t)$ extracted from $G(r)$ (points) and best fits to results (lines). (g,h) Structure factor obtained from simulations scaled by the characteristic length scale (points). Best fits to high-$k$ decay (lines). }
   \label{fig2}
\end{figure}

To simulate the quench dynamics we numerically evolve the spin-1 Gross-Pitaevskii equations \cite{Kawaguchi2012R} 
with initial condition of a  polar condensate $\bm{\psi}=\sqrt{n_0}(0,1,0)$ that has vacuum noise added to Bogoliubov modes for $q_i=\infty$ according to the truncated Wigner prescription \cite{Barnett2011a}.  At $t=0$ the quadratic Zeeman energy is suddenly reduced to its final value $q_f$ and dynamically unstable modes, seeded by the vacuum noise, cause magnetisation to grow exponentially \cite{Saito2007a,Leslie2009a,Barnett2011a} with a characteristic time scale $t_s=\hbar/2|g_s|n_0$ [Figs.~\ref{fig2}(a), (b)] and characteristic domain size  $\xi_s=\hbar/\sqrt{2M|g_s|n_0}$.
 The magnetization growth saturates (at $t\sim10^2\,t_s$ in our simulations) towards a value that is somewhat reduced from the ground state value due to a small thermal component that develops after the quench. This component arises from the thermalisation of the excess energy of the polar state over the ferromagnetic phase. 
As the magnetisation saturates we observe that the domains begin to coarsen and when they are large compared to $\xi_s$ [e.g.~Figs.~\ref{fig1}(b), (c)], the coarsening dynamics becomes universal and independent of the microscopic details.

In the universal regime we find, in agreement with the scaling hypothesis, that correlation functions of the order parameter have no explicit time-dependence when expressed in units of a characteristic length scale $L(t)$ \cite{Bray1994a}. We examine the single-time correlation function
\begin{align}
G(\bm{r},t)=\frac{1}{A}\int d^2\bm{x}\,\langle\phi(\bm{x}+\bm{r},t)\cdot\phi(\bm{x},t)\rangle,
\end{align}
for scalar or vector order parameter $\phi$, where $A$ is the area of the system and $\langle \rangle$ denotes an ensemble average. We calculate this correlation function by averaging over an ensemble of 8 simulations that differ only by the random seeding. Our simulations are performed on grids with $1024\times1024$ (easy-axis) or $2048\times2048$ (easy-plane) points, using an adaptive step spectral method and we ensure that the simulations accurately conserve normalisation, energy and the $z$-component of magnetisation.  The long-time coarsening of the domains is revealed by the spreading of this correlation function, shown in the insets to Figs.~\ref{fig2}(c) and (d).  For the easy-axis phase we take $L(t)$ to be the distance where  $G(r,t)$ first drops to zero, which is a measure of the average domain size. Because the easy-plane phase breaks a continuous symmetry, the notion of a single domain is not well defined.
We therefore take $L(t)$ to be the distance across which $G(r,t)$ drops to $0.25G(0,t)$.
 Upon rescaling spatial coordinates by this length scale, the correlation function at different times collapse onto a single function $G(r,t)\rightarrow f\left(r/L(t)\right)$, as shown in Figs.~\ref{fig2}(c) and (d), thus confirming the universal coarsening behaviour.

 We determine the dynamic critical exponent for the quench to the easy-axis phase by fitting $t^{1/z}$ to our results for $L(t)$ [see Fig.~\ref{fig2}(e)]. This yields $1/z=0.68$ in agreement with the $t^{2/3}$  growth law for a binary fluid in the inertial hydrodynamic regime \cite{Furukawa1985a} (also see \cite{Hofmann2014a}). We have found similar values for $z$ (to within fitting errors) for simulations performed with $q_f/|g_s|n_0=\{-1.2,-1.8,-2.4\}$. The binary fluid universality class is also known as model H \cite{Hohenberg1977a,bray2003}.

For the easy-plane phase we determine the dynamic critical exponent by fitting our results for $L(t)$ to $(t/\ln (t/t_0))^{1/z}$ [see Fig.~\ref{fig2}(e)]. This form has been used to describe coarsening dynamics in the XY-model from an initial condition containing free vortices \cite{Bray2000a} (also see \cite{Yurke1993a}), reflecting the slow approach to the asymptotic regime through the annihilation of vortex-antivortex pairs. In our simulations we observe that a large number of polar-core vortices \cite{Kawaguchi2012R,Kudo2015a} emerge in the initial unstable dynamics following the quench, which then decay away as $1/L(t)^2$ through vortex-antivortex annihilation. From this analysis we obtain $z=1.04$, which is consistent with $z=1$ for model E in a two dimensional system \cite{halperin1974} \footnote{A $t^{0.79}$ growth also fits the data in Fig.~\ref{fig2}(f). Deviation between growth with and without a logarithmic correction only becomes apparent for  $t/t_s\gtrsim 10^4$, much longer than our simulations can investigate.}. We have found similar values for $z$ (to within fitting errors) for simulations performed with $q_f/|g_s|n_0=\{0.2,1.2,1.8\}$.
Model E describes a non-conserved planar ferromagnet dynamically coupled to a second conserved field \cite{Hohenberg1977a}. This fits our system well, where the second conserved field is $F_z$ \cite{Lamacraft2007a,Hohenberg1977a,Nam2011a,tauber2014}. Incorporating conservation of energy into model E gives model E$^\prime$, which also has $z=1$ \cite{halperin1976}.  In the non-dissipative dynamics of a 2D (scalar) superfluid a value of $z=1$ was also observed \cite{Damle1996a}.

It is also convenient to consider the order parameter structure factor, obtained by Fourier transforming the correlation function 
\begin{align}
S(\bm{k},t)=\int d^2\bm{r}\, G(\bm{r},t)e^{i\bm{k}\cdot\bm{r}}=L^2\hat{f}(kL(t)),
\end{align}
where the scaling form follows from setting $G(r,t)= f\left(r/L(t)\right)$, with $\hat{f}$ being the Fourier transform of $f$. The structure factor is useful for examining the small $r/L$ properties of the order parameter, which can reveal the structure of domain walls and topological defects in the system \cite{Bray1994a}. Results for the structure factor for the easy-axis and easy-plane quenches are shown in Fig.~\ref{fig2}(g) and (h), respectively.

For the easy-axis case we observe a ``knee'' in the structure factor at $kL\sim 1.3$ followed by a ``Porod tail'' $S(k)\sim k^{-3}$ for $L> k^{-1}\gg \xi_s$ that indicates the presence of sharp domain walls \cite{Bray1994a}. For small $r/L$, the probability of two points a distance $r$ apart belonging to opposite domains is $r/L$, so that $G(r,t)\sim1-2r/L$. This linear dependence on small $r$ leads to the Porod law of $k^{-(d+1)}$ decay for a $d$-dimensional system with domain walls.

We also observe a Porod tail for the easy-plane case, but with a non-integer exponent, $S\sim k^{-2.45}$. We interpret this non-integer Porod tail as arising from the domains having a fractal surface structure \cite{bale1984,Schaefer1986a}. 
For domains in a $d$-dimensional system having surface fractal dimension $d_s$ a $k^{-2d+d_s}$ tail emerges in the structure factor    \cite{Oh1999a}, reducing to the usual Porod law for the smooth surface case $d_s=d-1$. 
Thus our results in Fig.~\ref{fig2}(h) suggest a surface fractal dimension of $d_s\approx 1.5$. To provide further evidence for this result, we determine a box-counting dimension for the domain boundaries. We bin the easy-plane order parameter into discrete domains based on the spin direction and perform a box counting algorithm on the boundaries of these domains over an order of magnitude of box sizes. This yields a box counting dimension of $D\approx 1.5-1.6$. For comparison, we have also applied this analysis to the easy-axis domain boundaries and extracted the box counting dimension of $D=1.0$. Possible physical implications of the fractal structure we observe includes diffusion limited aggregation \cite{halsey2000}, or Schramm (stochastic)-Loewner evolution and the associated conformal invariance \cite{cardy2005}. We note that the Porod tail in the easy-plane case is not accounted for by topological defects (vortices), which would result in a $k^{-4}$ tail \cite{Bray1991a,Bray1994a,Rojas1999a}.

A distinguishing feature of our system over more traditional models where coarsening dynamics has been observed is that our system has a firm microscopic foundation governed by conservative Hamiltonian evolution. It is therefore appealing to explore the long-time microscopic details of our system dynamics. In particular, it is of interest to consider the role of vortices in the coarsening dynamics. We note that recently an exact analytic treatment of the vortex dynamics in the XY-model provided further insights into their role in coarsening dynamics \cite{Forrester2013a}.
Previous work has shown that the Landau damping rates for the spin-wave excitations is a slow and ineffective thermalization mechanism in the post-quench dynamics \cite{Barnett2011a} but did not consider the role of spin vortices.

\begin{figure}[htbp] 
   \centering
   \includegraphics[width=3.5in]{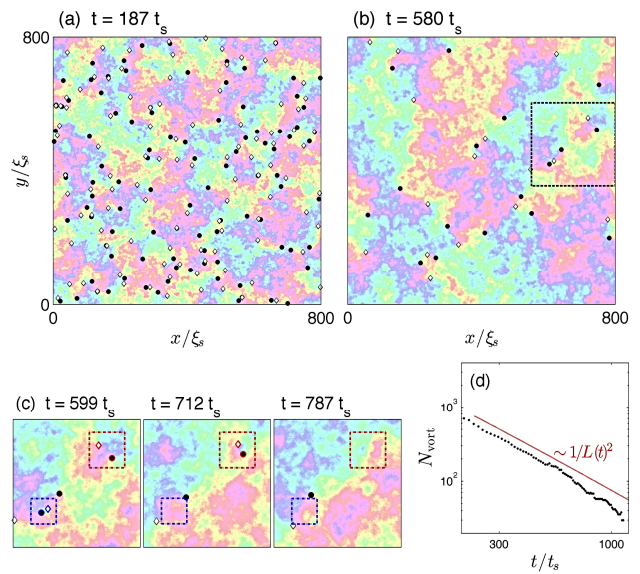} \\
   \vspace*{-0.25cm}
   \caption{Polar-core vortices with $\kappa\!=\!1$ ($\bullet$) and $\kappa\!=\!-1$  ($\diamondsuit$)  in the (a) early stages and (b) later stages of the easy-plane coarsening dynamics in a quadrant of the full simulation. Transverse magnetization indicated as in Fig.~\ref{fig1}(c), but with saturation reduced to make vortices clear.  (c) Evolution of the spatial region indicated with a dashed box in (b). The dashed boxes in (c) identify vortex-antivortex pairs that annihilate during the dynamics. (d) Total number of polar-core spin vortices  ($N_{\mathrm{vort}}$) as a function of time from a single quench simulation, demonstrating that the vortex density is proportional to $L(t)^{-2}$. Other parameters as in Fig.~\ref{fig1}(c).}
   \label{fig3}
\end{figure}

Following the easy-plane quench we identify the decay of singly charged polar-core vortices. The state of a polar-core vortex is $\bm{\psi}\sim\left(\sin\beta e^{-i\theta},\sqrt{2}\cos\beta,\sin\beta e^{i\theta}\right)^T$ where far from the vortex core $\cos\beta=\sqrt{(1+q/2|g_s|n)/2}$ \cite{Kawaguchi2012R}. The magnetisation lies in-plane with angle $\theta$ that rotates by $2\pi\kappa$ ($\kappa\in\mathbb{Z}$) around the vortex centre, giving rise to a spin current but no mass current. At the vortex centre the particle density concentrates in the $\psi_0$ component (hence  ``polar-core'').  While spin-1 condensates can support other vortices that combine mass and spin currents (e.g.~Mermin-Ho vortices) \cite{Kawaguchi2012R}, we only observe polar-core vortices of charge $\kappa=\pm 1$ (higher values of $\kappa$ are unstable). The vortices are indicated in Fig.~\ref{fig3}, revealing the decrease in vortex density as the coarsening progresses. This occurs as ($\kappa=1$) vortices and ($\kappa=-1$) anti-vortices are drawn together and annihilate, leading to domain annealing [see Fig.~\ref{fig3}(c)]. The quantitative relationship between the vortex decay and the coarsening is revealed in Fig.~\ref{fig3}(d). We also note that in the early stages of coarsening soliton like domain walls are observed  as notches in the magnitude of the transverse magnetisation. These decay due to snake-like instability \cite{Saito2007b} that produces a (polar-core) vortex anti-vortex pair.  
A model for the interaction of spin vortices in a ferromagnetic condensate was proposed in Ref.~\cite{Turner2009a}, but here spin-waves appear to affect the dynamics.
Thus a better understanding of the interaction of spin-waves with vortices is of interest, where it is possible that spin-waves provide an effective thermal field for the vortices during coarsening.

In summary, our  results for the Hamiltonian evolution of a spin-1 condensate quenched into a ferromagnetic phase reveals a wealth of universal dynamics that can be controllably explored in experiments.
 Importantly,  varying the final Zeeman energy  changes the order parameter symmetry, hence whether topological defects are supported, and also whether the order parameter is conserved during the dynamics: all crucial aspects of coarsening.  The first steps towards coarsening dynamics in the regime we consider have been made in experiments \cite{Sadler2006a,Leslie2009a,Vengalattore2010a}, and will be aided by recent developments of homogeneous trapping for cold gases \cite{Chomaz2015a,Navon2015a}, which is advantageous for studying critical phenomena. An exciting aspect in experiments arises from the capability to observe vortices \cite{Sadler2006a,Weiler2008a,Seo2015a} and potentially track  \textit{in situ}  domain dynamics \cite{Wilson2015a}.
Dipole-dipole interactions will have a role in the dynamics of some spinor gases, although these can be eliminated from the dynamics, e.g.~using radio-frequency pulses \cite{Vengalattore2008a}.
Initial condition dynamics are known to be important in XY coarsening dynamics. The zero temperature quench we consider here only leads to the production of polar-core vortices, while Mermin-Ho vortices are expected in the case of a quench from a sufficiently high temperature initial condition and may change the coarsening dynamics \cite{Kudo2015a}. Recently a suitable microscopic framework for simulating thermal dynamics of spinor condensates has been developed \cite{Bradley2014a}.

We acknowledge useful discussions with J.~Hofmann, Y.~Kawaguchi, K.~Kudo, M.~Reeves, J.~Brand, B.~Baeumer, X.-Yu, A.~Fetter, and C.~Chianca.
We gratefully acknowledge support by the Marsden Fund of the Royal Society of New Zealand (contract number UOO1220).

\end{document}